\documentclass[11pt,twocolumn]{article}

\usepackage[english]{babel}
\usepackage[utf8]{inputenc}
\usepackage[T1]{fontenc}
\usepackage{lmodern}
\usepackage{microtype}
\usepackage[
  letterpaper,
  top=0.85in,
  bottom=0.85in,
  left=0.7in,
  right=0.7in,
  footskip=0.4in,
  columnsep=0.3in
]{geometry}
\usepackage{titlesec}
\usepackage{fancyhdr}
\usepackage[font=small,labelfont=bf,labelsep=period]{caption}
\usepackage{graphicx}
\usepackage{amsmath,amssymb}
\usepackage{booktabs}
\usepackage{array}
\usepackage{multirow}
\usepackage{xcolor}
\usepackage{colortbl}
\usepackage{hyperref}
\usepackage[numbers,sort&compress]{natbib}
\usepackage{enumitem}
\usepackage{float}
\usepackage{tikz}
\usepackage{pgfplots}
\pgfplotsset{compat=1.18}
\usetikzlibrary{arrows.meta,positioning,shapes.geometric,calc,patterns,fit}

\hypersetup{
  colorlinks=true,
  linkcolor=blue!50!black,
  citecolor=blue!50!black,
  urlcolor=blue!50!black,
}

\definecolor{tableheader}{HTML}{1F4E79}
\definecolor{tableheaderalt}{HTML}{A6192E}
\definecolor{lightgray}{HTML}{F6F6F6}

\titleformat*{\section}{\large\bfseries}
\titleformat*{\subsection}{\normalsize\bfseries}
\titleformat*{\subsubsection}{\normalsize\itshape}
\titlespacing*{\section}{0pt}{10pt}{5pt}
\titlespacing*{\subsection}{0pt}{8pt}{3pt}
\titlespacing*{\subsubsection}{0pt}{6pt}{2pt}

\fancypagestyle{plain}{
  \fancyhf{}
  \fancyfoot[C]{\small\itshape Preprint. Under review.}
  \fancyhead[R]{\small Page \thepage}
  
}

\setlist{nosep,leftmargin=1.2em}

\begin{document}

\twocolumn[
\begin{@twocolumnfalse}
\begin{center}
{\LARGE\bfseries Agentic AI in the Software Development Lifecycle:\\[4pt]
\large Architecture, Empirical Evidence, and the Reshaping of Software Engineering\par}
\vspace{12pt}
{\large \textbf{Happy Bhati}\\[2pt]
Northeastern University $\cdot$ Boston, Massachusetts, USA\\[1pt]
\texttt{bhati.h@northeastern.edu}}
\vspace{10pt}
\end{center}

\begin{center}
\begin{minipage}{0.92\textwidth}
\small
\textbf{Abstract.}
The arrival of large language models (LLMs) capable of multi-step reasoning, tool use,
and long-horizon planning has produced a qualitative shift in software engineering.
Where earlier code-completion tools such as GitHub Copilot operated at the granularity
of a line or function, modern \emph{agentic} systems---Claude Code, OpenAI Codex CLI,
Google Jules, Devin, OpenHands, SWE-agent, MetaGPT, ChatDev, and DeepMind's
AlphaEvolve---operate at the granularity of a repository, a feature, or an algorithm.
We synthesize work from Anthropic, OpenAI, Google DeepMind, Microsoft Research,
Princeton, Stanford, and the broader academic community to characterize this transition.
We propose a six-layer reference architecture for agentic software engineering systems,
contrast a traditional Software Development Lifecycle (SDLC) with an emerging
\emph{Agentic SDLC} (A-SDLC), and consolidate empirical evidence on performance
(a rise from 1.96\% to 78.4\% on SWE-bench Verified between October 2023 and April 2026),
productivity (13.6\%--55.8\% time savings across controlled studies), and labor-market
impact (49\% of jobs sampled by Anthropic in 2026 saw AI used for at least a quarter of
their tasks). We argue that the central object of inquiry has shifted from
\emph{code generation} to \emph{delegated execution under human supervision}, and we
identify five open problems---evaluation, governance, technical debt, skill
redistribution, and the economics of attention---that will determine whether the agentic
transition is net-positive for the discipline. This paper is intended as both a synthesis
for practitioners and a research agenda for the field.

\vspace{4pt}
\end{minipage}
\end{center}
\vspace{10pt}
\end{@twocolumnfalse}
]

\section{Introduction}

The release of GitHub Copilot in 2021 marked the first mainstream encounter most
software engineers had with a generative-AI tool integrated into their daily workflow
\cite{peng2023impact}. Copilot, like its successors Tabnine, Codeium, and Amazon
CodeWhisperer, operated as an in-editor \emph{completion} system: the developer wrote,
the model suggested, the developer accepted or rejected. The cognitive contract was
clear---the human remained the engineer; the model was an autocomplete with judgment.

Five years later, that contract has been substantially renegotiated. Anthropic's Claude
Code, OpenAI's Codex CLI, Google's Jules, Cognition's Devin, and the open-source
OpenHands and SWE-agent projects do not complete code so much as \emph{perform engineering
work}: they read a repository, formulate a plan across multiple files, execute shell
commands, run tests, observe failures, revise their approach, and deliver a committed
change \cite{anthropic2025claude37,yang2024sweagent,anthropic2026claudecode,wang2025openhands}.
Anthropic reports that the majority of code written internally is now produced by Claude
Code \cite{anthropic2026claudecode}. On Princeton's SWE-bench Verified benchmark---
designed to test resolution of real GitHub issues drawn from popular open-source
repositories---state-of-the-art systems have moved from 1.96\% in October 2023 to roughly
78\% by spring 2026 \cite{jimenez2024swebench,openai2024verified,anthropic2026opus47}.

This shift is not merely quantitative. It changes \emph{what software engineers do}
during a working day, \emph{what artifacts} the SDLC produces, and \emph{which skills}
command a wage premium. It also raises governance questions---about determinism,
auditability, security, and intellectual property---that the discipline has not yet
fully metabolized.

\paragraph{Contributions.}
This paper makes four contributions. First, in \S\ref{sec:background} we synthesize
the principal lines of work on agentic coding from major industrial labs and academic groups.
Second, in \S\ref{sec:architecture} we propose a six-layer
reference architecture that organizes the design space of agentic software engineering
systems. Third, in \S\ref{sec:asdlc} we contrast the traditional SDLC with an Agentic
SDLC (A-SDLC). Fourth, in \S\ref{sec:evidence} and \S\ref{sec:comparative}
we consolidate empirical evidence on performance and labor-market effects, and in
\S\ref{sec:open} identify five open problems that constitute the field's
near-term research agenda.

\section{Background and Related Work}
\label{sec:background}

\subsection{From Code Completion to Agency}

The earliest LLM-based coding tools were trained on large corpora of public source code
and evaluated on benchmarks like HumanEval and MBPP, both of which test the ability to
synthesize a single function from a docstring \cite{chen2021codex,austin2021mbpp}.
These benchmarks saturated quickly: by 2024, frontier models exceeded 90\% pass@1 on
HumanEval. The saturation of single-function evaluation prompted Jimenez et al.\ at
Princeton to introduce \emph{SWE-bench} in October 2023 \cite{jimenez2024swebench},
a benchmark of 2{,}294 GitHub issues drawn from twelve mature Python repositories.
SWE-bench requires a system to navigate a real codebase, locate the relevant files,
write a patch that resolves the issue, and pass the project's hidden test suite. The
original report found that no system, retrieval-augmented or otherwise, could resolve
more than $\sim$2\% of issues.

The performance ceiling broke not because models became larger, but because the
\emph{scaffolding} around them changed. SWE-agent \cite{yang2024sweagent}, introduced by
the same Princeton group at NeurIPS 2024, demonstrated that a custom \emph{agent--computer
interface} (ACI)---structured commands for navigating, editing, and testing---lifted
resolution rates to 12.5\% even with the same underlying model. This insight, that
\emph{interface design for the agent matters as much as model capability}, has organized
the field since.

\subsection{Industrial Frontier Systems}

\paragraph{Anthropic.} Anthropic released Claude Code as a research preview alongside
Claude 3.7 Sonnet in early 2025 \cite{anthropic2025claude37,anthropic2025claude4}.
Claude Code operates at project granularity: it reads the full codebase, plans changes
across files, executes shell commands, runs tests, and iterates. Anthropic reports that
Claude Opus 4.7 leads on SWE-bench Verified at 78.4\%
\cite{anthropic2026opus47}. The product positions itself as a \emph{delegation}
interface---engineers define goals and review results rather than guiding each step.

\paragraph{OpenAI.} OpenAI's contribution has been bracket-shaped. In 2021, the original
Codex paper introduced HumanEval \cite{chen2021codex}. In 2024, OpenAI collaborated with
the SWE-bench authors to release SWE-bench Verified, a 489-task subset filtered for
solvability and clarity \cite{openai2024verified}. By 2026, GPT-5.4-Codex powers a Codex
CLI that competes directly with Claude Code, achieving high scores on Terminal-Bench 2.0
and SWE-bench Pro \cite{openai2026codex}.

\paragraph{Google DeepMind.} DeepMind's AlphaEvolve
\cite{novikov2025alphaevolve,deepmind2025alphaevolve}, unveiled in May 2025, represents
a different conception of agentic coding. Rather than resolving GitHub issues, AlphaEvolve
uses an evolutionary loop---an ensemble of Gemini Flash and Pro models proposes program
variants, an automated evaluator scores them, and the best survive into the next
generation. AlphaEvolve has discovered a new $4\times 4$ complex matrix multiplication
algorithm using 48 scalar multiplications (improving on Strassen's 1969 result), advanced
20\% of fifty open mathematical problems, recovered 0.7\% of compute across Google's data
centers, and produced a 23\% speedup of the FlashAttention kernel that trains Gemini
itself \cite{novikov2025alphaevolve}.

\paragraph{Cognition / Devin.} Cognition's Devin, demonstrated in March 2024, was the
first commercial product to popularize the framing of ``AI software engineer''
\cite{cognition2024devin}. Devin runs in a sandboxed cloud VM with a browser, terminal,
and editor, and is dispatched tasks via Slack, Jira, or its web UI. Although Devin's
headline SWE-bench numbers have been overtaken by frontier-model agents, the product
established the pattern that competitors---commercial and open-source alike---now follow.

\paragraph{Microsoft / GitHub.} Microsoft Research conducted some of the earliest
rigorous productivity studies on Copilot \cite{peng2023impact,dohmke2023seachange}.
In a controlled experiment, developers using Copilot completed an HTTP server task
55.8\% faster than the control group \cite{peng2023impact}; a follow-up analysis of
934{,}533 Copilot users found a $\sim$30\% suggestion acceptance rate and projected
\$1.5T in cumulative GDP impact by 2030 \cite{dohmke2023seachange}. Microsoft's
Azure-based agentic SDLC reference architecture \cite{parida2025sdlc} formalizes
specialized phase-agents coordinated by a core orchestrator.

\subsection{Academic and Open-Source Work}

The academic literature on agentic software engineering has expanded rapidly. MetaGPT
\cite{hong2024metagpt} and ChatDev \cite{qian2024chatdev} both encode software
development as a multi-agent process: MetaGPT instantiates Product Manager, Architect,
Engineer, and QA roles operating under standardized operating procedures, while ChatDev
models a virtual software company in which agents communicate through a chat-chain.
AgileCoder \cite{nguyen2024agilecoder} extends these ideas with sprints and a dynamic
code-dependency graph. OpenHands (formerly OpenDevin) \cite{wang2025openhands} is the
leading open-source generalist platform, providing a Docker-sandboxed execution
environment and supporting multiple LLM backends. AutoCodeRover \cite{zhang2024autocoderover}
and HyperAgent \cite{phan2024hyperagent} specialize in repository-level program repair.
Recent surveys \cite{wang2025agentssurvey,liu2025surveybench,sapkota2025agenticsurvey}
organize this work along axes of planning, memory, tool augmentation, and self-reflection.

\begin{figure}[t]
  \centering
  \begin{tikzpicture}
    \begin{axis}[
      width=\columnwidth,
      height=4.8cm,
      xlabel={\small Timeline},
      ylabel={\small SWE-bench Verified (\%)},
      xmin=0, xmax=7,
      ymin=0, ymax=90,
      xtick={0.5,1.5,2.5,3.5,4.5,5.5,6.5},
      xticklabels={Oct'23, Mid'24, Late'24, Ear'25, Mid'25, Late'25, Apr'26},
      xticklabel style={font=\tiny, rotate=25, anchor=east},
      yticklabel style={font=\tiny},
      ytick={0,20,40,60,80},
      grid=major,
      grid style={gray!20},
      legend style={at={(0.03,0.97)}, anchor=north west, font=\tiny,
                    draw=gray!40, fill=white, fill opacity=0.9},
      every axis plot/.append style={thick},
    ]
    \addplot[blue!70!black, mark=*, mark size=1.8] coordinates {
      (0.5,1.96) (1.5,12.5) (2.5,33.2) (3.5,49) (4.5,62.3) (5.5,72.7) (6.5,78.4)
    };
    \addlegendentry{Agentic systems}
    \addplot[red!60!black, dashed, mark=triangle*, mark size=1.8] coordinates {
      (0.5,1.96) (1.5,8) (2.5,14) (3.5,18) (4.5,19) (5.5,20) (6.5,20)
    };
    \addlegendentry{Non-agentic (RAG)}
    \node[font=\tiny, blue!70!black, anchor=south west] at (axis cs:6.2,78.4) {78.4\%};
    \node[font=\tiny, red!60!black, anchor=north west] at (axis cs:6.2,20) {$\sim$20\%};
    \end{axis}
  \end{tikzpicture}
  \caption{Issue-resolution rate on SWE-bench Verified, October 2023 to April 2026.
  Agentic systems (solid) compounded rapid gains; non-agentic RAG-based LLMs (dashed)
  plateaued near 20\%. Data compiled from
  \cite{yang2024sweagent,jimenez2024swebench,openai2024verified,anthropic2026claudecode,anthropic2026opus47,openai2026codex}.}
  \label{fig:swebench}
\end{figure}
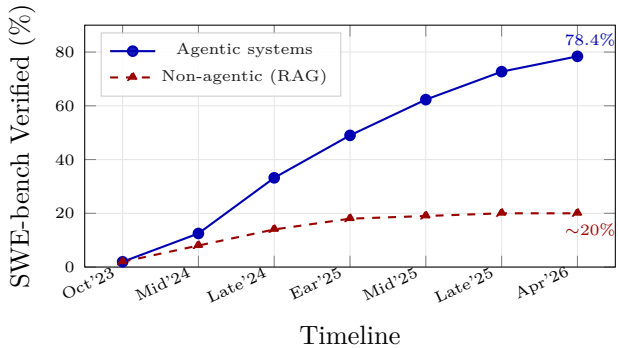

\section{A Reference Architecture for Agentic Software Engineering}
\label{sec:architecture}

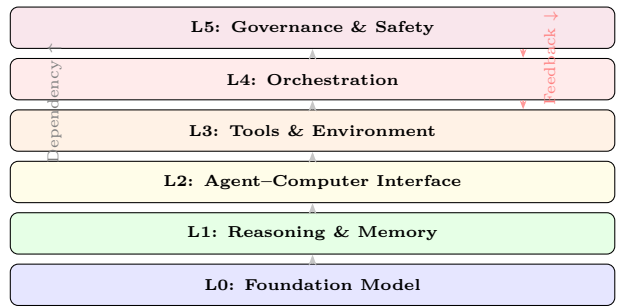
\begin{figure}[t]
  \centering
  \begin{tikzpicture}[
    layer/.style={draw, rounded corners=3pt, minimum width=\columnwidth-0.6cm, minimum height=0.55cm,
                  font=\tiny, align=center},
    arr/.style={-{Stealth[length=1.5mm]}, semithick, gray!45},
    node distance=0.12cm
  ]
  \node[layer, fill=purple!10] (L5) {\textbf{L5: Governance \& Safety}};
  \node[layer, fill=red!8, below=of L5] (L4) {\textbf{L4: Orchestration}};
  \node[layer, fill=orange!10, below=of L4] (L3) {\textbf{L3: Tools \& Environment}};
  \node[layer, fill=yellow!10, below=of L3] (L2) {\textbf{L2: Agent--Computer Interface}};
  \node[layer, fill=green!10, below=of L2] (L1) {\textbf{L1: Reasoning \& Memory}};
  \node[layer, fill=blue!10, below=of L1] (L0) {\textbf{L0: Foundation Model}};
  \draw[arr] (L0) -- (L1); \draw[arr] (L1) -- (L2);
  \draw[arr] (L2) -- (L3); \draw[arr] (L3) -- (L4); \draw[arr] (L4) -- (L5);
  \draw[-{Stealth[length=1.2mm]}, semithick, red!40, dashed]
    ([xshift=2.8cm]L5.south) -- ([xshift=2.8cm]L4.north);
  \draw[-{Stealth[length=1.2mm]}, semithick, red!40, dashed]
    ([xshift=2.8cm]L4.south) -- ([xshift=2.8cm]L3.north);
  \node[font=\tiny, gray, rotate=90, anchor=south] at (-3.2,-1.0) {Dependency $\uparrow$};
  \node[font=\tiny, red!50, rotate=90, anchor=south] at (3.4,-0.4) {Feedback $\downarrow$};
  \end{tikzpicture}
  \caption{Six-layer reference architecture. Dependency runs upward; feedback flows
  downward (dashed red).}
  \label{fig:architecture}
\end{figure}

Across heterogeneous systems we observe a recurring stratification, summarized in
Figure~\ref{fig:architecture}. We organize it as six layers (L0--L5):

\paragraph{L0: Foundation Model.}
The base LLM provides the system's reasoning and code-generation capacity. Claude Opus 4.7,
GPT-5.4, and Gemini 3.1 Pro currently anchor the frontier; open-weights models (Llama,
DeepSeek, Qwen-Coder) populate self-hosted deployments.

\paragraph{L1: Reasoning, Memory, Self-Reflection.}
The cognitive scaffolding: chain-of-thought and ReAct-style planning
\cite{wei2022cot,yao2023react}, short- and long-term memory mechanisms (including
``memory files'' produced natively by Claude Opus 4 \cite{anthropic2025claude4}), and
self-critique loops. Self-reflection largely accounts
for the gap between zero-shot model accuracy and agentic resolution rates.

\paragraph{L2: Agent--Computer Interface (ACI).}
The ACI \cite{yang2024sweagent} translates between an LLM's
text-token output and concrete operations on a real computer. The Princeton
group's core finding is that ACI design quality is, empirically, as
important as model size.

\paragraph{L3: Tools and Environment.}
Filesystem and editor operations, shell
and process management, web browsing, test runners and compilers, and
integration with version control and CI/CD.

\paragraph{L4: Orchestration.}
Two dominant patterns coexist. Single-agent loops (SWE-agent
\cite{yang2024sweagent}, Claude Code \cite{anthropic2026claudecode}) maintain one
cognitive locus. Multi-agent systems---MetaGPT,
ChatDev, AgileCoder, AgentMesh
\cite{hong2024metagpt,qian2024chatdev,nguyen2024agilecoder,khanzadeh2025agentmesh}---
decompose work across role-specialized agents.

\paragraph{L5: Governance and Safety.}
Permission boundaries, sandboxing, audit logs, and policies for sensitive
operations. We argue in \S\ref{sec:open} that L5 is currently the least mature layer and is
rapidly becoming the bottleneck on enterprise deployment.

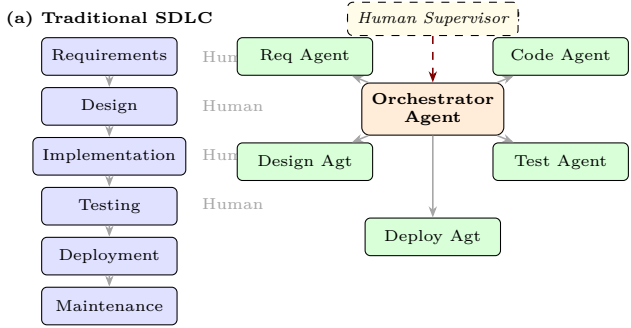
\begin{figure}[t]
  \centering
  \begin{tikzpicture}[
    phase/.style={draw, rounded corners=2pt, minimum width=1.8cm, minimum height=0.5cm,
                  font=\tiny, align=center, fill=#1},
    arr/.style={-{Stealth[length=1.5mm]}, semithick, gray!70},
    lbl/.style={font=\tiny\bfseries},
    node distance=0.15cm
  ]
  \node[lbl] at (1.2,3.2) {(a) Traditional SDLC};
  \node[phase=blue!12] (r1) at (1.2,2.7) {Requirements};
  \node[phase=blue!12, below=of r1] (d1) {Design};
  \node[phase=blue!12, below=of d1] (i1) {Implementation};
  \node[phase=blue!12, below=of i1] (t1) {Testing};
  \node[phase=blue!12, below=of t1] (dp1) {Deployment};
  \node[phase=blue!12, below=of dp1] (m1) {Maintenance};
  \draw[arr] (r1) -- (d1); \draw[arr] (d1) -- (i1);
  \draw[arr] (i1) -- (t1); \draw[arr] (t1) -- (dp1); \draw[arr] (dp1) -- (m1);
  \foreach \n/\y in {r1/2.7,d1/2.05,i1/1.4,t1/0.75} {
    \node[font=\tiny, gray!70, anchor=west] at (2.3,\y) {Human};
  }

  \node[lbl] at (5.5,3.2) {(b) Agentic SDLC};
  \node[draw, rounded corners=3pt, fill=orange!15, minimum width=1.6cm, minimum height=0.6cm,
        font=\tiny\bfseries, align=center] (orch) at (5.5,2.0) {Orchestrator\\Agent};
  \node[phase=green!15] (ra) at (3.8,2.7) {Req Agent};
  \node[phase=green!15] (da) at (3.8,1.3) {Design Agt};
  \node[phase=green!15] (ca) at (7.2,2.7) {Code Agent};
  \node[phase=green!15] (ta) at (7.2,1.3) {Test Agent};
  \node[phase=green!15] (dpa) at (5.5,0.3) {Deploy Agt};
  \draw[arr] (orch) -- (ra); \draw[arr] (orch) -- (da);
  \draw[arr] (orch) -- (ca); \draw[arr] (orch) -- (ta);
  \draw[arr] (orch) -- (dpa);
  \node[draw, dashed, rounded corners=2pt, fill=yellow!10, minimum width=1.6cm,
        minimum height=0.4cm, font=\tiny\itshape] (human) at (5.5,3.2) {Human Supervisor};
  \draw[-{Stealth[length=1.5mm]}, semithick, red!50!black, dashed] (human) -- (orch);
  \end{tikzpicture}
  \caption{Traditional SDLC (a) vs.\ Agentic SDLC (b). In the A-SDLC, an orchestrator
  coordinates specialized sub-agents while the human supervises at intent, review, and
  approval gates.}
  \label{fig:sdlc}
\end{figure}

\section{From SDLC to Agentic SDLC}
\label{sec:asdlc}

The classical SDLC---whether waterfall, iterative, or Agile---presupposes that behavior
is fully specified at build time and validated before release \cite{epam2026adlc}.
Agentic systems violate this assumption in two ways. First, the systems themselves are
stochastic: prompt drift, context truncation, and model updates produce non-deterministic
behavior under nominally identical inputs. Second, the \emph{development process}
becomes stochastic too: an agent may take a different path through the same task on
different runs.

We follow recent industry frameworks \cite{parida2025sdlc,epam2026adlc,codebridge2026adlc}
in distinguishing an \emph{Agentic SDLC} (A-SDLC) from the augmented SDLC merely
accelerated by AI assistants. Figure~\ref{fig:sdlc} illustrates the contrast, and
Table~\ref{tab:sdlc} maps the phase-by-phase translation.

\begin{table}[t]
  \centering
  \scriptsize
  \renewcommand{\arraystretch}{1.15}
  \setlength{\tabcolsep}{3pt}
  \begin{tabular}{p{1.2cm} p{2.8cm} p{2.8cm}}
    \toprule
    \rowcolor{tableheader}
    \textcolor{white}{\textbf{Phase}} &
    \textcolor{white}{\textbf{Traditional SDLC}} &
    \textcolor{white}{\textbf{Agentic SDLC}}\\
    \midrule
    Require\-ments & Analyst writes specs; stakeholder review &
                   Intent specification; agent drafts spec\\
    \rowcolor{lightgray}
    Design & Architect produces diagrams, ADRs &
             Agents propose \& critique; human selects\\
    Implement. & Engineers write code file by file &
                     Coding agents execute plan; human reviews\\
    \rowcolor{lightgray}
    Testing & QA writes/runs tests &
              Testing agents generate suites; sandbox\\
    Deploy & Manual or pipeline promotion &
                 CI agent gates; human approves prod\\
    \rowcolor{lightgray}
    Maintain. & On-call, tickets, hotfixes &
                  Monitor agents triage; repair agents patch\\
    KPI & Cycle time, defect rate &
                  Accept rate, escalation quality\\
    \bottomrule
  \end{tabular}
  \caption{Phase-by-phase comparison of traditional and Agentic SDLC.}
  \label{tab:sdlc}
\end{table}

Three structural differences are worth emphasizing. First, the unit of work shrinks:
instead of estimating two-week sprints, teams scope work in tasks that an agent can
complete in minutes to hours, with human review at the boundary. Second, the developer's
role shifts from \emph{producing} to \emph{orchestrating, reviewing, and directing}---
closer to the role of a senior engineer or tech-lead than to that of an individual
contributor \cite{anthropic2026claudecode}. Third, behavioral metrics---agent acceptance
rate, escalation quality, supervision burden---displace, but do not entirely replace,
process metrics like cycle time and defect rate \cite{epam2026adlc,codebridge2026adlc}.

\section{Empirical Evidence}
\label{sec:evidence}

\subsection{Capability}

Figure~\ref{fig:swebench} plots SWE-bench Verified resolution rates for the dominant
agentic system at each timepoint between October 2023 and April 2026. The trajectory is
approximately logistic: from 1.96\% (RAG baseline \cite{jimenez2024swebench})
to 12.5\% (SWE-agent \cite{yang2024sweagent}) to 33.2\% (Anthropic's first
scaffold around Sonnet 3.5 \cite{anthropic2024swebench}) to 49\% (Claude 3.5 Sonnet new
\cite{anthropic2025claude4}) to 62.3\% (Claude 3.7 Sonnet + Claude Code
\cite{anthropic2026claudecode}) to 72.7\% (Claude Sonnet 4 \cite{anthropic2025claude4})
to 78.4\% (Claude Opus 4.7 \cite{anthropic2026opus47}). Frontier non-agentic systems
have plateaued near 20\%, confirming that the gain is dominated by scaffolding rather than
raw model capability.

\begin{figure}[t]
  \centering
  \begin{tikzpicture}
    \begin{axis}[
      width=\columnwidth,
      height=5cm,
      xlabel={\small Autonomy Level},
      ylabel={\small Capability (\%)},
      xmin=0, xmax=10,
      ymin=0, ymax=90,
      xtick={1,3,5,7,9},
      xticklabels={Complete, Suggest, Guided, Semi-auto, Auto},
      xticklabel style={font=\tiny, rotate=10, anchor=east},
      yticklabel style={font=\tiny},
      ytick={0,20,40,60,80},
      grid=major,
      grid style={gray!15},
      legend style={at={(0.03,0.97)}, anchor=north west, font=\tiny,
                    draw=gray!40, fill=white, fill opacity=0.9},
    ]
    \addplot[only marks, mark=*, mark size=3, blue!70!black, fill=blue!30]
      coordinates {(8.5,78.4) (8,73) (7.5,70)};
    \addlegendentry{Frontier lab}
    \addplot[only marks, mark=square*, mark size=2.5, red!60!black, fill=red!25]
      coordinates {(9,14)};
    \addlegendentry{Autonomy-first}
    \addplot[only marks, mark=triangle*, mark size=2.5, green!50!black, fill=green!25]
      coordinates {(6,12.5) (6.5,40)};
    \addlegendentry{Open-source}
    \addplot[only marks, mark=diamond*, mark size=2.5, orange!70!black, fill=orange!25]
      coordinates {(3,50)};
    \addlegendentry{Editor-integrated}
    \node[font=\tiny, anchor=south west] at (axis cs:8.5,79) {Claude Code};
    \node[font=\tiny, anchor=south east] at (axis cs:7.9,74) {Codex CLI};
    \node[font=\tiny, anchor=south west] at (axis cs:7.5,71) {Jules};
    \node[font=\tiny, anchor=west] at (axis cs:9.2,14) {Devin};
    \node[font=\tiny, anchor=south east] at (axis cs:5.8,13) {SWE-agent};
    \node[font=\tiny, anchor=south] at (axis cs:6.5,41) {OpenHands};
    \node[font=\tiny, anchor=east] at (axis cs:2.8,51) {Copilot};
    \addplot[dashed, gray!60, no marks, domain=2:9, samples=30] {10*exp(0.24*x)};
    \end{axis}
  \end{tikzpicture}
  \caption{The capability--autonomy frontier, 2021--2026. Frontier-lab stacks (blue)
  define the upper-right envelope; open-source systems (green) trade capability for
  accessibility; Devin (red) prioritizes autonomy. Dashed curve: approximate frontier.}
  \label{fig:frontier}
\end{figure}
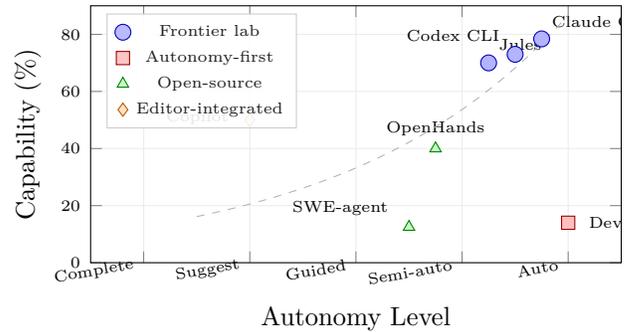

Figure~\ref{fig:frontier} places major systems on a two-dimensional
\emph{capability--autonomy} plane. The frontier is currently anchored by
frontier-lab agentic stacks: Claude Code, GPT-5.4 + Codex CLI, and Gemini 3.1 Pro + Jules.
AlphaEvolve sits off the SWE-bench frontier because it targets algorithmic discovery
\cite{novikov2025alphaevolve}.

\subsection{Productivity}

Productivity evidence falls into three tiers. \emph{Controlled experiments} have reported
the largest effects: Peng et al.\ \cite{peng2023impact} found a 55.8\% completion-time
reduction on a JavaScript HTTP-server task; Brandebusemeyer et al.\
\cite{brandebusemeyer2026} found that moderate Copilot use improved efficiency, while
excessive use eroded the benefit. \emph{Field experiments}: Cui et al.\
\cite{dohmke2023seachange} observed 12.92\%--21.83\% more pull requests per week at
Microsoft and 7.51\%--8.69\% at Accenture. \emph{Longitudinal team studies}
found that team performance and perceived efficiency
rise even when commit metrics stay flat \cite{economicindex2026}.

These numbers should be read alongside two cautionary findings. First, recent work argues
that AI-assisted code can \emph{increase} long-term technical debt \cite{bauer2025debt}.
Second, productivity gains are unevenly distributed: experienced Claude users are substantially
more successful than newcomers \cite{economicindex2026}, suggesting a learning curve that compounds
the existing skill distribution.

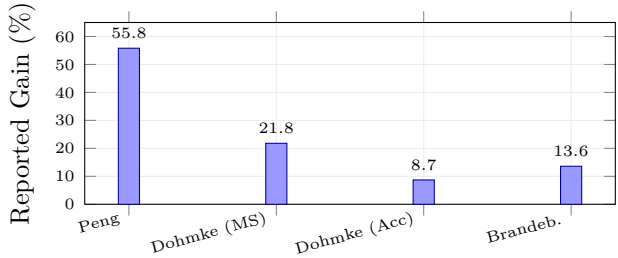
\begin{figure}[t]
  \centering
  \begin{tikzpicture}
    \begin{axis}[
      ybar,
      width=\columnwidth,
      height=4cm,
      bar width=8pt,
      ylabel={\small Reported Gain (\%)},
      symbolic x coords={Peng, Dohmke (MS), Dohmke (Acc), Brandeb.},
      xtick=data,
      xticklabel style={font=\tiny, rotate=15, anchor=east},
      yticklabel style={font=\tiny},
      ymin=0, ymax=65,
      ytick={0,10,20,30,40,50,60},
      grid=major,
      grid style={gray!15},
      nodes near coords,
      nodes near coords style={font=\tiny},
      every node near coord/.append style={anchor=south},
    ]
    \addplot[fill=blue!40, draw=blue!60!black] coordinates {
      (Peng,55.8) (Dohmke (MS),21.8) (Dohmke (Acc),8.7) (Brandeb.,13.6)
    };
    \end{axis}
  \end{tikzpicture}
  \caption{Productivity gains across four studies of AI-assisted coding
  \cite{peng2023impact,dohmke2023seachange,brandebusemeyer2026,economicindex2026}.
  Controlled experiments report the largest effects.}
  \label{fig:productivity}
\end{figure}

\subsection{Adoption and Labor-Market Signals}

Anthropic's quarterly Economic Index reports \cite{anthropic2025index,anthropic2025primitives,economicindex2026}
are the most systematic public source of usage data. As of February 2026, computer and
mathematical tasks account for $\sim$35\% of Claude.ai conversations and nearly half of
API traffic. 49\% of jobs in Anthropic's sample saw Claude used for at least a quarter
of their tasks (up from 36\% in January 2025). However, when usage is weighted by an
\emph{effective coverage} measure, software developers
are \emph{relatively less affected} than na\"ive task-coverage would suggest---a
counterintuitive finding driven by the difficulty of the tasks developers
actually delegate.

Anthropic's labor-market analysis \cite{anthropic2026labor} finds no clear unemployment
signal in high-exposure occupations as of early 2026, but does find that hiring of
workers aged 22--25 into the most exposed roles has slowed by $\sim$14\% relative to a
counterfactual. Among 81{,}000 respondents to a separate survey \cite{anthropic202681k},
roughly one-fifth of workers in AI-exposed jobs voiced concern about economic
displacement; software engineers were among the most concerned.

\section{Comparative Analysis of Frontier Programs}
\label{sec:comparative}

We summarize the strategic posture of each major program in Table~\ref{tab:programs}.
Three patterns are notable. \emph{Anthropic} has placed primary emphasis on the
\emph{delegation} use case---long-horizon agentic coding with explicit human approval
gates. \emph{OpenAI} has converged on a similar product surface (Codex CLI) and competes
closely on benchmarks \cite{openai2026codex}. \emph{Google DeepMind} has pursued a distinctive
evolutionary path with AlphaEvolve \cite{novikov2025alphaevolve}. \emph{Microsoft / GitHub}
occupies the broadest user base. \emph{Princeton/Stanford} open-source
work on SWE-agent and SWE-bench underpins the evaluation
infrastructure \cite{yang2024sweagent,jimenez2024swebench,openai2024verified}.

\begin{table}[t]
  \centering
  \scriptsize
  \renewcommand{\arraystretch}{1.15}
  \setlength{\tabcolsep}{2pt}
  \begin{tabular}{p{1.3cm} p{1.8cm} p{2.6cm} p{0.8cm}}
    \toprule
    \rowcolor{tableheaderalt}
    \textcolor{white}{\textbf{Program}} &
    \textcolor{white}{\textbf{Flagship}} &
    \textcolor{white}{\textbf{Emphasis}} &
    \textcolor{white}{\textbf{SWE-b}}\\
    \midrule
    Anthropic & Claude Code (Opus 4.7) & Delegation; memory; safety & $\sim$78\%\\
    \rowcolor{lightgray}
    OpenAI & Codex CLI (GPT-5.4) & General agentic capability & $\sim$73\%\\
    DeepMind & AlphaEvolve + Jules & Evolutionary discovery & $\sim$70\%\\
    \rowcolor{lightgray}
    MS/GitHub & Copilot, Azure & Enterprise breadth & $\sim$50\%\\
    Cognition & Devin & Sandboxed AI SWE & $\sim$14\%\\
    \rowcolor{lightgray}
    Princeton & SWE-agent/bench & Benchmarks \& ACI & 12.5\%\\
    Academia & MetaGPT, ChatDev & Multi-agent SOPs & n/a\\
    \bottomrule
  \end{tabular}
  \caption{Frontier agentic-coding programs, c.~April 2026. SWE-bench Verified figures
  are approximate
  \cite{yang2024sweagent,anthropic2026claudecode,anthropic2026opus47,anthropic2025claude4,openai2026codex}.}
  \label{tab:programs}
\end{table}

Despite differing strategies, convergence is visible. All major programs now expose:
(i) a CLI- or IDE-resident agent with shell, file, and test-runner access;
(ii) human-in-the-loop approval at high-impact actions; (iii) some form of long-term
memory; (iv) parallel/multi-agent execution for non-trivial tasks; and (v) a strong story
about safety and audit.

\section{Open Problems and Research Agenda}
\label{sec:open}

We close by identifying five open problems we believe will dominate the next
two-to-three years of work.

\subsection{Evaluation beyond SWE-bench}
SWE-bench has been an indispensable forcing function, but it has
limitations: it is dominated by Python, focuses on bug-fixing, and provides ground-truth
tests that real-world tasks lack \cite{yang2024multimodal,zhao2025swecompass}.
SWE-bench Multimodal \cite{yang2024multimodal}, SWE-bench Pro, Terminal-Bench, and
SWE-Compass \cite{openai2026codex,yang2024multimodal,gao2025swebenchcl} are extending
coverage to JavaScript, multi-language, multimodal, and production-aligned tasks.
Research is needed on benchmarks that capture long-horizon delegation,
collaboration with human reviewers, faithfulness to stated
intent, and the absence of ``reward hacking'' behaviors \cite{anthropic2025claude4}.

\subsection{Governance, Safety, and Audit}
The 2025 AI Agent Index \cite{casper2026agentindex} documents a proliferation of deployed
agentic systems alongside a marked deficit in governance documentation.
The ADLC framework \cite{epam2026adlc,codebridge2026adlc}
proposes \emph{human--agent responsibility mapping}---an explicit allocation of authority
levels and approval gates---as a non-optional design step. Operationalizing this idea
in a way auditors can verify is an open problem.

\subsection{The Technical-Debt Hypothesis}
Bauer et al.\ \cite{bauer2025debt} argue that AI-assisted programming may decrease
productivity for experienced developers by inflating maintenance burden. Agents are biased toward
producing \emph{more} code (because production is cheap) and toward \emph{local} fixes
(because global redesign is expensive in tokens). Long-term studies of repository health
under sustained agent contribution are urgently needed.

\subsection{Skill Redistribution}
Anthropic's data \cite{economicindex2026} suggests an emerging two-track market:
experienced engineers capture compounding gains;
newcomers use agents to \emph{do} the work and underperform. Educational research
that explicitly trains \emph{orchestration skills}---decomposition, prompting,
review, and judgment about when not to delegate---will be needed to keep the pipeline
open.

\subsection{The Economics of Attention}
If an agent can produce ten plausible patches per hour, the rate-limiting resource is
human review. Research on tooling for high-throughput review---automated diff
summarization, behavioral test generation, principled sampling of agent traces---is a
natural successor research program to the agent programs of 2023--2026.

\section{Conclusion}

Software engineering is in the middle of a discontinuity. In 2021, the question was
whether language models could plausibly autocomplete a function; by 2026, frontier
agentic systems resolve roughly four out of five real GitHub issues drawn from mature
repositories, and a non-trivial fraction of the code at major AI labs is
written by such systems \cite{anthropic2026claudecode}.

Our reading is cautiously optimistic. Agentic systems are real productivity tools, not
toys; the empirical evidence on time saved and tasks delegated is substantial, if uneven.
But the frame of \emph{code generation} is the wrong frame. The right frame is
\emph{delegated execution under human supervision}---and the long-run winners will be those who invest earliest in the process,
governance, and skills that delegation requires. The research agenda we have sketched
is, in the end, a research agenda for keeping humans firmly in the loop while letting
the machines do the typing.

\section*{Acknowledgments}

We thank the open-source maintainers of SWE-bench, SWE-agent, and OpenHands, and the
public documentation teams at Anthropic, OpenAI, Google DeepMind, Microsoft Research,
and Cognition, whose published reports made this synthesis possible.

\bibliographystyle{plainnat}
\bibliography{references}

\end{document}